\documentclass[12pt,a4paper]{article}
  \usepackage[margin=5mm,a5paper]{geometry}
\usepackage{mathptmx,nicefrac}      
\usepackage{microtype, amsmath, amssymb, url}
\usepackage{cleveref}
\usepackage{caption,subcaption,graphicx,bm}
\usepackage[numbers]{natbib}
\usepackage[left]{lineno}
\title{Ligand mediated adhesive mechanics of two deformed spheres}

\author{
Sarthok Sircar\thanks{Corresponding author \protect\url{mailto:sarthok.sircar@adelaide.edu.au}} 
\and
Andrei Kotousov\thanks{School of Mechanical Engineering\protect\url{mailto:andrei.kotousov@adelaide.edu.au}}
\and
Giang Nguyen\thanks{School of Civil Engineering\protect\url{mailto:g.nguyen@adelaide.edu.au}}
\and
Anthony J. Roberts\thanks{School of Mathematical Sciences, University of Adelaide, South Australia~5005, Australia. 
\protect\url{mailto:anthony.roberts@adelaide.edu.au}}
}

\renewcommand{\vec}[1]{\text{\boldmath$#1$}}
\newcommand{\ben}{\begin{equation}}
\newcommand{\een}{\end{equation}}

\begin{document}
\maketitle

\begin{abstract}
A self-consistent model is developed to investigate attachment / detachment kinetics of two soft, deformable microspheres with irregular surface and coated with flexible binding ligands. The model highlights how the microscale binding kinetics of these ligands as well as the attractive/repulsive potential of the charged surface affects the static deformed configuration of the spheres. It is shown that in the limit of smooth, neutral charged surface (i.e., the Debye length, $\kappa \rightarrow \infty $), interacting via elastic binders (i.e., the stiffness coefficient, $\lambda \rightarrow 0$) the adhesion mechanics approaches the regime of application of the JKR theory, and in this particular limit, the contact radius scales with the particle radius, according to the scaling law, $R_c\propto R^{\nicefrac{2}{3}}$. We show that adhesion dominates in larger particles with highly charged surface and with resilient binders. Normal stress distribution within the contact area fluctuates with the binder stiffness coefficient, from a maximum at the center to a maximum at the periphery of the region. Surface heterogeneities result in a diminished adhesion with a distinct reduction in the pull off force, larger separation gap, weaker normal stress and limited area of adhesion. These results are in aggrement with the published experimental findings.
\end{abstract}


\section{Introduction} \label{sec:intro}
Biological adhesion is couched in a very different vocabulary from the mechanical adhesion theory, although there is a tremendous overlap of applications~\citep{vonByern2010}. Examples include binding of bacterial clusters to medical implants or host cell surfaces during infection~\citep{Zhu2000}, cancer cell metastasis~\citep{Moss2000}, coalescence of medical gels with functionalized particles or micro-bubbles for targeted drug delivery~\citep{Sircar2013} and more recent applications in Micro-Electro-Mechanical Systems (MEMS) and nanotechnology~\citep{Zhang2013}. Molecular bioadhesion is commonly mediated by specific ligand interactions, e.g., the ligand-mediated surface adhesion is an important case in the experimental studies of the \textsc{p}-selectin/\textsc{psgl}-1 catch bond interactions of leukocytes (a roughly spherical particle) with and without fluid flow~\citep{Marshall2003}. The adhesive properties of biological surfaces connected by multiple independent tethers are also presently inspiring the development of novel adhesives mimicking the remarkable properties of beetle and gecko feet~\citep{Varenberg2007}. Several other applications as well as {\it in vivo} and {\it in silico} studies of biological adhesion are listed in~\citep{Lauffenburger1993,Springer1995,Hammer1996,Jones1996,Zhu2000}. However, the models and the experiments listed in these references fail to capture the coupled effects of ligand mediated adhesion interrelated with the nonlinear mechanics of surface deformation. Therefore the motivation of this article is to develop a unified theory and approach that can capture these coupled effects.

The theoretical modelling of the adhesion of deformed, charged, spherical surfaces presents significant challenges. Early work utilised simple Hertzian deformation characteristics assuming that the spheres were elastic indentors~\citep{Hertz1882} but adhesion was unsustainable in that approach since the attractive, pull off force and the area of contact is zero at detachment. Subsequently, two major theories were developed within the elastic framework: the Johnson-Kendall-Roberts (JKR)~\citep{JKR1971} and the Derjaguin-Muller-Toporov (DMT)~\citep{DMT1975} theories. The JKR theory assumes that the adhesive forces act within the contact region but are absent outside of it, resulting in infinite normal stress either at the center or at the edge of the contact zone. The DMT theory deploys cohesive surface forces outside the contact zone, while retaining the Hertzian force-deformation characteristics in the core. This results in the adhesive stress being zero inside the contact area and finite outside it. Unlike the DMT case, JKR theory predicts a non-zero adhesion area at the critical separation gap at pull-off. These two apparently contradictory theories were described as opposite extremes of a parameter by Tabor~\citep{Tabor1977}. 

The adhesive forces are composed of numerous physical processes all of which determines the fate of the binding surfaces including ligand-receptor binding kinetics~\citep{Dembo1988}, surface deformation and the related mechanical stresses due to the elastic forces~\citep{Hodges2002}, excluded volume effects and paramagnetism~\citep{Forest2006}, and multiscale short range interactions~\citep{Sircar2013}. Consequently, many detailed kinetic models have successfully, yet independently, described these physical features imperative in the adhesion-fragmentation processes. Schwarz {et al.}~\citep{Korn2006} and more recently Mahadevan {et al.}~\citep{Mani2012} studied the biological adhesion between the ligand coated wall and a rigid sphere moving in a shear flow. A similar model by Seifert {et al.}~\citep{Bihr2012} described the membrane adhesion via Langevin simulations. On the contrary, the macro-scale phase-field models describe the geometry of aggregates as a continuum mass of extracellular polymeric substance and predict the stability of the anisotropic structures in a flowing medium~\citep{Cogan2004}. However, a bridge between the receptor bond kinetics and the deformation of the elastic surface, detailing the several, multi-scale, interrelated phenomena in the adhesion process is still missing~\citep{King2005a}, but is now addressed here.

The aim of this article is to develop and investigate a unified, multi-scale model (with spatial variation at the nano-micro level) of the adhesion kinetics of the ligand smeared spheres integrated with the mechanics of solid, micro-particles with soft, deformable material and with heterogeneous surfaces. Our model is important case from an experimental perspective. For example, consider the experiments by Sokurenko et al.~\citep{Sokurenko1997, Sokurenko1998} which reveal the catch bond interactions of FimH proteins attached to the surface of E. coli in stagnant conditions. In the next section, we present a comprehensive description of a self-consistent model, including the bond mechanics (\S \ref{subsec:BK}), long range interactions via charged surfaces (\S \ref{subsec:LRI}), micro-scale binding forces on the particle surface (\S \ref{subsec:MF}), the calculation of the adhesion area of deformed irregular surface (\S \ref{subsec:AA}), the normal stress distribution, pull off force and adhesion energy (\S \ref{subsec:NS}). Section~\ref{sec:results} highlights the numerical outcomes of the case studies of the static, deformed configuration of the ligand coated spheres in a select range of material parameters, which concludes (\S \ref{sec:end}) with a brief discussion of the implication of these results and the focus of our future directions.

\section{Mathematical model: binder kinetics, long range interactions, forces and contact area} \label{sec:model}
\begin{figure}
\centering
\includegraphics[scale=0.4]{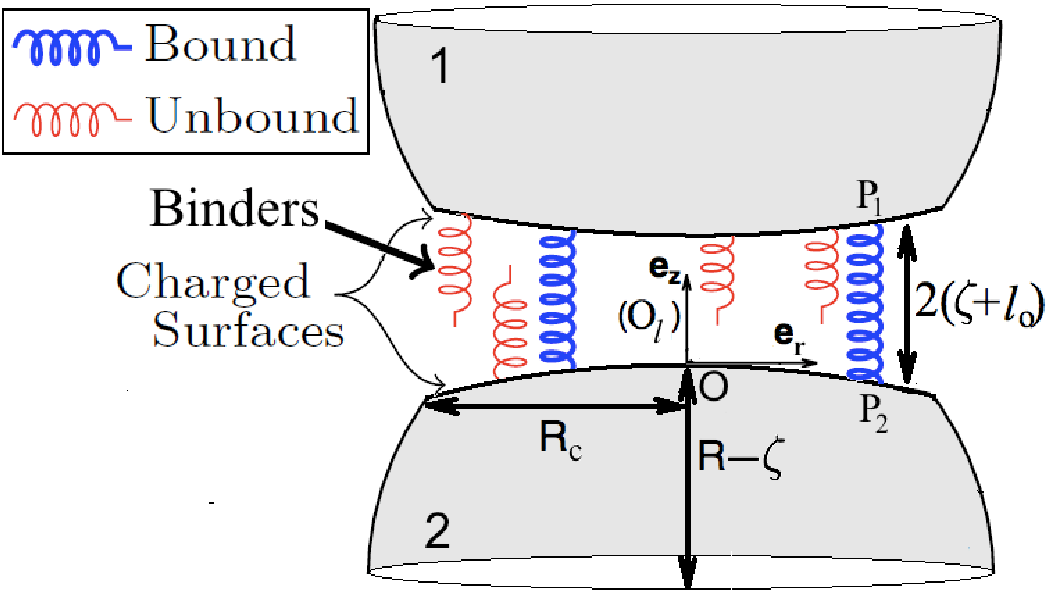}
\caption{A schematic of two static, deformed spheres coated with an adhesive, modelled via the binding ligands (\S\ref{subsec:BK}). The symbol~$(R_m)$ denotes the local frame of reference, with the origin $O_m$ fixed on the surface of sphere~$2$ along the line joining the centre of the sphere.}\label{fig:Fig1}
\end{figure}

This section describes the adhesion mechanics of two static, heterogeneously charged, deformable, solid,  equal sized, spherical microparticles (figure~\ref{fig:Fig1}). The spheres deform normal to the surface and adhere within well-defined disc-like patches. Adhesion is achieved via the attachment/detachment of the binders (i.e., polymer strands with sticky heads) normal to the adhering surface~\citep{Sircar2013,Sircar2014}. A heterogeneously charged surface is characterized by a fraction, $\Theta$, which denotes the real area of adhesion (details given in \S \ref{subsec:AA}).

Further, the binder kinetics is assumed to be independent of the salt concentration (i.e. the spring stiffness is independent of the Debye length and the zeta potential). This implies that we are neglecting the electro-viscous stresses~\citep{Tabatabaei2010}. Due to the relatively large micron-size scale, the binding kinetics of these spheres are significantly different from the core-shell nano-crystal interactions, which are applicable at much smaller scales~\citep{Duval2008}. The next few subsections presents detailed aspect of this model.

\subsection{Binder kinetics} \label{subsec:BK}
Figure~\ref{fig:Fig1} illustrates the model of the interfacial attachment of two deformed spheres, with an identical size of radius~\(R\) (when undeformed). To simplify the visualisation of the system, consider a local frame,~$(O_{\it l})$, with origin~$O$, fixed on the surface of the sphere $2$ at a point equidistant from the edge of the separation gap. In this frame of reference, the unit vectors~\(\vec e_r\), \(\vec e_z\) denote the directions, radial and normal to the surface, respectively. Let \(\zeta(r)\) denote the net, axis-symmetric deformation of either of the spheres (i.e. the deformation of the sphere surface minus the rigid translation in the transverse direction) and $r$ is the distance from the centre, $O$, along the radial direction. Thus, the separation gap between the spheres is \(2D(r)=2(\zeta(r)+{\it l}_0)\). The binders are idealized as linear Hookean springs with stiffness $\lambda$ and \({\it l}_0\)~the mean rest length. Define $A_{\text{Tot}} g(r)\, d A$ as the number of bonds that are attached within the surface~$d A$, where $A_{\text{Tot}}$~is the bond density per unit area. Thus, the total number of bonds formed is $\int_{A_c} A_{\text{Tot}} g(r)\, d A$\,, where~$A_c$ is the area of adhesion (refer \S \ref{subsec:AA}). In established research on colloids, the function~$g$ is synonymous with the term \emph{sticking probability}~\citep{Somasundaran2005}.

The forward and reverse reaction rates for the ligand binding are written as Boltzmann distributions, allowing highly stretched bonds to be readily broken by thermal energy fluctuations as well as due to the interaction potential, $W(D)$, arising from charged surfaces (see \S \ref{subsec:LRI}). With these degrees of freedom, the bond attachment\slash detachment rates are 
\begin{align}
K_{\text{on}} (r) &= K_{\text{on,eq}} \exp \left[ \frac{- \lambda_s(D(r)-{l}_0)^2 + W(D(r))}{2{k}_B T} \right ], \nonumber \\
K_{\text{off}} (r) &= K_{\text{off,eq}} \exp \left[ \frac{ (\lambda - \lambda_s)(D(r)-{l}_0)^2 + W(D(r)) }{2{k}_B T} \right], \label{eqn:reaction_rates}
\end{align}
where ${k}_B$ is the Boltzmann constant, \(T\) is the temperature, $\lambda_s$~is the spring constant of the transition state (see Dembo {\it et al.}~\citep{Dembo1988}) used to distinguish the catch bonds ($\lambda < \lambda_s$) from slip bonds ($\lambda > \lambda_s$). In the limit of small binding affinity and abundant ligands on the binding surface (i.e., $K_{eq}=\nicefrac{A_{\text{Tot}} K_{\text{on, eq}}}{K_{\text{off, eq}}} \ll 1$), the bond ligand density evolves in accordance with~\citep{Dembo1988, Hodges2002, Reboux2008}
\ben
\frac{d g}{d t} = A_{\text{Tot}} K_{\text{on}} - K_{\text{off}} g\,, \quad g = 0 \quad  \text{for }   x \ge R_c \,,
\label{eq:coll_fac}
\een
However, in the case of static adhesion, assuming that the attachment/detachment rates of the flocs are sufficiently rapid so that the non-equilibrium binding kinetics can be ignored (i.e., set $\frac{d g}{d t}$ = 0 in Eqn.(\ref{eq:coll_fac})), the evolution of the ligand bond density reduces to
\ben
g(D) = K_{\text{eq}} \exp\left[ \frac{- \lambda(D(r)-{l}_0)^2}{2{k}_B T} \right ]
\een
We remark that $g \in [0, K_{\text{eq}}]$ and in the limit of small binding affinity ($K_{\text{eq}} \ll 1$), this value cannot exceed 1. In further description of the model we denote \(D(r) \equiv D\), without loss of generalisation. 

\subsection{Long range interactions}\label{subsec:LRI}
Derjaguin, Landau, Verwey and Overbeek (DLVO) theory is utilized to describe the interaction between the charged cell surfaces via a surface potential~$W(D)$ (in Eqn.~\eqref{eqn:reaction_rates}).
The current study incorporates the effects of Coulombic repulsion and Van der Waals attraction only.
Other interactions including hydration effects, hydrophobic attraction, short range steric repulsion, and polymer bridging, which are absent in the length scales of our interest, are neglected~\citep{Gregory2006}.
For two charged spheres, with an identical size of radius~$R$, the potential due to the Coulombic force in a gap of size~$2D$ is 
\ben
W_{\text{C}}(D) = 2\pi \varepsilon_0 \varepsilon \psi^2 R e^{-2\kappa D} ,
\een
where $\kappa$ is the Debye length, $\varepsilon$ and~$\varepsilon_0$ are the dielectric constant of vacuum and the medium, respectively, and $\psi$~is the average \emph{zeta potential} or the electric potential of the diffuse cloud of charged counterions.
The potential due to the Van der Waal forces for these spherical particles within the same gap width is 
\ben
W_{\text{VW}}(D) = -\frac{A R}{24 D},
\een  
where \(A\) is the Hamaker constant measuring the van der Waal `two-body' pair-wise interaction for macroscopic spherical objects.

\subsection{Microscale forces} \label{subsec:MF}
We non-dimensionalize the length scales with respect to the undisturbed radius of the cell,~\(R\), and introduce the following dimensionless variables denoted by bar ( $\bar{}$ ),
\ben
D=\bar{D}R, \quad K_{\text{on}} = \bar{K}_{\text{on}} K_{\text{on,eq}}\,, \quad K_{\text{off}} = \bar{K}_{\text{off}} K_{\text{off,eq}}\,, \quad g = \bar{g} K_{\text{eq}}. \label{eqn:ND}
\een
Further, let us introduce the following non-dimensional parameters,
\ben
\gamma = \frac{\lambda {l}^2_0}{{k}_B T}\,, \quad \bar{\lambda}_s = \frac{\lambda_s}{\lambda}\,, \quad \epsilon = \frac{{l}_0}{R} \,.
\label{eqn:param}
\een
Then, the non-dimensional form of the reaction rates, Eqn.~\eqref{eqn:reaction_rates} and the bond density evolution Eqn.~\eqref{eq:coll_fac}, are
\begin{align}
\bar{K}_{\text{on}} &= \exp \Big[ -\bar{\lambda}_s \frac{{\gamma}}{2\epsilon^2} (\bar{D} - \epsilon)^2 + \bar{W}(D) \Big] \,, \nonumber
\\
\bar{K}_{\text{off}} &= \exp \Big[ (1-\bar{\lambda}_s)\frac{{\gamma}}{2\epsilon^2}(\bar{D} - \epsilon)^2 + \bar{W}(D) \Big] , \label{eqn:Konoff}
\\
\bar{g}(r) &= e^{-\frac{\gamma}{\epsilon^2}(\bar{D}-\epsilon)^2}, \label{eqn:g*}
\end{align}
where $\bar{W}(D)={W(D)}/({2{k_B}T})$. To evaluate the microscale forces, consider one individual bond formed between two points on the surface of the spheres (say, point P$_1$ on sphere 1 and point P$_2$ on sphere 2, figure~\ref{fig:Fig1}). The instantaneous force it exerts on the two spheres has three components normal to the surface: an extensional force related to bond stretching given by Hooke's law, ${\vec f}_E = \lambda(D - {l}_0) {\vec e}_z$\,; force due to Coulombic repulsion, ${\vec f}_C = \nabla W_C(D) {\vec e}_z$ and van der Waal attraction, ${\vec f}_{VW} = \nabla W_{VW}(D) {\vec e}_z$\,. The operator, $\nabla$, denotes the derivative with respect to $D$. Thus, the microscale point force due to one bound ligand is~\citep{Reboux2008},
\ben
{\vec f}(r)={\bf f}_E+{\bf f}_C+{\bf f}_{VW}. \label{eqn:microForce}
\een

%
%
%
\subsection{Contact area of heterogeneously charged spheres} \label{subsec:AA}
Previous experimental studies have revealed that surface heterogeneities severely hinders particle adhesion~\citep{Santore2015}. An irregular surface can have variations ranging from those which are larger in size than the local separation gap, to the nanometer scale heterogeneities which has a scale much smaller than the local asperity curvature. Thus the study of real contact is a multiscale problem~\citep{Duffadar2008}. This work considers nanometer scale heterogeneities, much smaller than the radii of the micron sized particles, but comparable in size to the separation distance. Thus, the heterogeneities can be assumed to be uniformly, spatially distributed. We characterize a heterogeneous surface by the fraction, $\Theta$, such that area of adhesion of the two spheres is $A_c=\Theta \pi R^2_c$\,, where $R_c$~is the contact radius (and $\Theta=1.0$ coincides with smooth surfaces, figure~\ref{fig:Fig1}). In further description we develop the theory for frictionless, smooth interacting surfaces, which can be readily extended for the heterogeneous case (\S \ref{subsec:SH}).

Although the contacting surfaces never really touch, we treat the elastic surface deflection of the spheres as if they were plain, solid particles~\citep{Hertz1882}. According to the half-space theory, for a given axisymmetric normal point force, ${\vec f}(r)$, the normal surface displacement, $\zeta(r)$, is given by (Eqn.~3.22b in~\citep{Johnson1985})
\ben
\zeta(r) = \left(\frac{1-\nu^2}{\pi E}\right) f_0 \int_0^{R_c} \frac{4 t}{t+r} \bar{\vec f}(r) \bar{g}(r) K(k) dt, \label{eqn:zetaDeflection}
\een
where $\nu$ and $E$ are the Poisson ratio and the Young modulus of the elastic spheres, respectively, and $f_0=\lambda {\it l}_0$. $K(k)$ is the complete elliptic integral of the first kind, and $k = \nicefrac{2\sqrt{tr}}{(t+r)}$. For a consistent solution, we require that the microscale forces arising from the ligand kinetics to vanish at the edge of the contact, i.e.,
\ben
\bar{\vec f}(\zeta_1)|_{r=R_c} = 0, \quad \frac{d \bar{\vec f}}{d \zeta}\Big |_{\zeta_1, r=R_c} > 0.\label{eqn:forceCont}
\een
%
The coupled system of Eqns.~(\ref{eqn:zetaDeflection}, \ref{eqn:forceCont}) is numerically solved to estimate the static equilibrium displacement, $\zeta_1(r)$, and the contact radius, $R_c$. The second constraint in Eqn.~(\ref{eqn:forceCont}) is imposed to distinguish the solution from the other equilibrium deformation, $\zeta_0(r)$ (e.g., shown in figure~\ref{fig:fig6}\subref{subfig:fig6a}), where
\ben
\bar{\vec f}(\zeta_0)|_{r=R_c} = 0, \quad \frac{d \bar{\vec f}}{d \zeta}\Big |_{\zeta_0, r=R_c} < 0.\label{eqn:zeta0}
\een

%
\subsection{Normal stresses} \label{subsec:NS}
The normal stresses produced by a concentrated normal force, ${\vec f}(r)$ acting within a circular area (of radius $R_c$) on the surface of an elastic half-space (Eqn. 3.20a and 3.20c in~\citep{Johnson1985}) are
\begin{align}
\sigma_r &= \left(\frac{1-2\nu}{2\pi}\right) \frac{f_0}{R^2} \int^{R_c}_{0} \int^{2\pi}_{0} \frac{\bar{\vec f}(t)\bar{g}(t)}{s^2(r,t)} t \,d \theta \, dt \nonumber
\\ 
& = (1-2\nu) \frac{f_0}{R^2} \int^{R_c}_{0} \frac{\bar{\vec f}(t)\bar{g}(t)}{|r^2-t^2|} t \, dt, \label{eqn:sigmaR}
\end{align}
along the radial direction, and
\begin{align}
\sigma_z &= \lim_{z \to 0} \frac{F_0}{R^2} \int^{R_c}_{0} \int^{2\pi}_{0} \frac{-3\bar{\vec f}(t)\bar{g}(t)}{2\pi}\frac{z^3}{(s^2(r,t)+z^2)^{5/2}} t \,d \theta \, dt \nonumber
\\
&= \lim_{z \to 0} \frac{27}{384} \frac{F_0}{R^2} \int^{R_c}_{0} \frac{z^3}{r^{5/2}t^{3/2}} {\bar{\vec f}(t)\bar{g}(t)} \, q\!\left(\frac{r^2+t^2+z^2}{t r}\right) \, dt, \label{eqn:sigmaZ}
\end{align}
along the direction normal to the surface. $F_0 = K_{eq}f_0$. The normal stress on the surface of the deforming spheres along the angular direction is, $\sigma_{\theta}=-\sigma_r$ (Eqn. 3.20b in~\citep{Johnson1985}), and therefore not described here. $s(r,t)=\sqrt{r^2+t^2-2tr\cos\theta}$ is the distance between the position of the application of the normal force, $t$, and the position where the deformation is evaluated, $r$, inside the area of contact (see figure 3.6 in~\citep{Johnson1985} for description). The function $q(x)$ is defined by
%
%
%
\ben
q(x) = p^{-5/2} \left[ 1- \frac{1}{5}\left(\frac{5}{2}\right)\frac{p}{x} \ldots + \left(-\frac{p}{x}\right)^n\frac{4!}{(n+3)!}{-\nicefrac{5}{2}\choose n} + \ldots\right],
\een
and $p=\frac{x}{2}-\sqrt{|\left(\frac{x}{2}\right)^2 -1|}$. We remark that the angular integration in Eqns.~(\ref{eqn:sigmaR},\ref{eqn:sigmaZ}) are solved analytically by calculating appropriate residues of singularities inside a unit circle in the complex plane~\citep{Ablowitz2003}. Further note that, depending on the material properties, the normal stresses are singular at selected points inside the region of adhesion (e.g., consider the case studies presented in figure~\ref{fig:fig4}\subref{subfig:fig4a}).

Next, define the total microscale force arising from all the bound ligands on the sphere surface, as
\ben
F(\zeta(r)) = -2\pi F_0 \int^{R_c}_0 \bar{g}(r) \bar{\vec f}(r) \bar{r} \, d \bar{r}, \label{eqn:Ftot}
\een
and the adhesive force as the total force within the range of equilibrium deformations where the microscale force is attractive
\ben
F_a(\zeta(r)) = \left\{ F(\zeta(r)) \mid \zeta_0(r)\le \zeta(r) \le \zeta_1(r) \right\}.
\label{eqn:Fa}
\een
The pull off force is defined as the maximum attractive force given by
\ben
F_{pull} = \min_{0 \leq r \leq R_c} \left\{ F_a(\zeta_c(r)) \mid \left( \nicefrac{d F}{d \zeta}\right)_{\zeta_c(r)}=0 \right\}, \label{eqn:Fpull}
\een
and the adhesion energy, $w$, is the work done against this adhesive force to cleave two identical spherical surface,
\ben
w = -F_0 R \int^{R_c}_0 \bar{r} d\bar{r} \int^{\zeta_1(r)}_{\zeta_0(r)} \bar{g}(r) \bar{f}(r) d \bar{\zeta}. \label{eqn:SurfEnergy}
\een
Finally, eqns.~(\ref{eqn:Konoff}--\ref{eqn:SurfEnergy}) represent the system of equations that fully describe the binding kinetics of two deformed, heterogeneously charged, static spheres. The next section describes the numerical results and exposits the biophysical implications of this system.

\section{Results and discussion} \label{sec:results}
The values of the material parameter used in our numerical calculations are listed in Table~\ref{tab:parameter}. The parameter values are chosen so that they closely replicate the static ligand-receptor kinetics of neutrophiles. For example, the \textsc{p}-selectine molecule extends about \(40\)\,nm from the endothelial cell membrane, so when combined with its ligand \textsc{psgl}-1 it is reasonable to take ${l}_0 \approx 100$\,nm as an estimate of the length of the unstressed bond~\citep{Shao1998}.
Typically, neutrophils have a size of $R \approx 1-10\,\mu$m which gives the length ratio $\epsilon \approx 0.1-0.01$ (equation~\eqref{eqn:ND}). Hochmuth~\citep{Shao1998} measured variations of up to eight orders of magnitude \emph{in vivo} for the values of the  microvillus stiffness,~$\lambda$, as well as the membrane tension of an undisturbed cell. Direct measurements of the parameters,~$A_{\text{Tot}}$, $K_{\text{on,eq}}$ and~$K_{\text{off,eq}}$ are scarce, although values in several thousands have been used in previous models~\citep{Hammer1996}. Fang~\citep{Fang2009} estimated the Young's modulus, $E$, and the poisson ratio, $\nu$, for a variety of hydrated polyacrylamide (PAA) gels with embedded impurities, which mimicks the elastic modulus of the soft material of the cells. The dielectric constant in vacuum is $\varepsilon_0 = 8.854 \times 10^{-12}$\,Farad\,m$^{-1}$, whereas the permittivity of water at temperature~$25^\circ$C is $\varepsilon=78.5$ (not to be confused with~$\epsilon$ which is a length ratio in Eqn.~\eqref{eqn:ND}). A zeta potential of $\psi=25$ mV is chosen which corresponds to the surface potential studies in~\citep{Gregory2006}, Chap-3. The Boltzmann factor is taken as {\it k}$_B$T=$4 \times 10^{-21}$J. The Hamaker constant measuring the macroscopic Van der Waal sphere-sphere interaction is fixed at $A=2.44${\it k}$_B$T~\citep{Gregory2006}. In the next two sections, we address the static equilibrium of the attachment of two smooth, deformed, charged spheres interacting via flexible tethers.
\begin{table}
\centering
\caption{Parameters common to all numerical results and used in  studies of the system of equations~(\eqref{eqn:Konoff}--\eqref{eqn:SurfEnergy}).}\label{tab:parameter}
\begin{equation*}
\begin{array}{ccccc}
\hline
\text{Parameter} & \text{Value} & \text{Units} & \text{Source}\\
\hline
A_{\text{Tot}} & 10^9 & \text{m}^{-2} & \text{\citep{Hammer1996}} \\
K_{\text{on, eq}} & 10^2 & \text{s}^{-1} & \text{\citep{Hammer1996}} \\
K_{\text{off, eq}} & 10^{14} & \text{s}^{-1} & \text{\citep{Hammer1996}} \\
\lambda & 10^{-8}-1 & \text{N\,m}^{-1} & \text{\citep{Mani2012}} \\
{\it l}_0 & 10^{-7} & \text{m} & \text{\citep{Shao1998}} \\
R & (1-10) \times 10^{-6} & \text{m} & \text{\citep{Shao1998}} \\
E & 5 & \text{kPa} & \text{\citep{Fang2009}} \\
\nu & 0.48 & - & \text{\citep{Fang2009}} \\
\hline
\end{array}
\end{equation*}
\end{table}

\subsection{Experimental validation} \label{subsec:EV}
The self-consistent system of nonlinear integral Eqns.~\eqref{eqn:zetaDeflection}, \eqref{eqn:forceCont} are numerically solved using the adaptive Lobatto quadrature (via Matlab function \verb|quadl|) to calculate the unknowns, $\zeta$, the net deformation and $R_c$, the radius of contact for static adhesion. We do not wish to study the effect of the material properties of the particle on the ligand induced adhesion kinetics and therefore the Young's modulus is fixed at $E=5$ kPa (refer table~\ref{tab:parameter}).

We note that in the limit of smooth, neutrally charged spheres (i.e., the Debye length, $\kappa \rightarrow \infty $) and in the case of perfect adhesion via highly elastic binders (i.e., the spring stiffness coefficient, $\lambda \rightarrow 0$), the adhesive force (Eqn.~\eqref{eqn:microForce}) reduces to that of a purely attractive van der Waal case (i.e., ${\vec f}(r) \rightarrow {\vec f}_{VW}$) and the adhesion energy, $w \rightarrow \infty$ (since the equilibrium deformations, $\zeta_0 \rightarrow 0$, $\zeta_1 \rightarrow \infty$, Eqn.~(\ref{eqn:forceCont}, \ref{eqn:zeta0}, \ref{eqn:SurfEnergy})). Eqn.~\eqref{eqn:zetaDeflection} can be evaluated analytically, yielding the net deformation $\zeta = \mathcal{O}(R^{1/3})$. This is the case of a strong surface adhesion and significant deformation and in this regime the JKR theory applies~\citep{JKR1971}, which predicts the relationship between the contact radius and the particle size as follows
\ben
R_c^3 = \frac{R_h}{K}\left\{ F_{\textrm{pull}}+3\pi R_h w + \sqrt{6\pi R_hwF_{\textrm{pull}}+(3\pi R_hw)^2} \right\}, \label{eqn:JKRradius}
\een
where $R_h = \nicefrac{R}{2}$ is the harmonic mean of the radii of the two spheres. In the limit of strong adhesion (i.e., $w \rightarrow \infty$), the JKR theory (Eqn.~\eqref{eqn:JKRradius}) predicts the scaling law: $R_c \propto R^{2/3}$.

Another approach to analyze this case would be to estimate the Tabor parameter~\citep{Tabor1977}, $\mu=\frac{Rw^2}{G^2\zeta^3_0}$, which is the ratio of the elastic displacement of the surface at the point of separation (pull-off) to the effective range of surface forces characterized by the equilibrium gap, $\zeta_0$, at which adhesive forces vanish. $G=\nicefrac{E}{1-\nu^2}$ is the effective elastic modulus of the microspheres. In the limiting case of purely attractive van der Waal force, the adhesion energy $w \rightarrow \infty$, and the equilibrium spacing $\zeta_0 \rightarrow 0$, which implies that the Tabor parameter $\mu \rightarrow \infty$, which is the region of application of the JKR theory.

\begin{figure}
\centering
\begin{subfigure}{0.49\textwidth}
\includegraphics[width=\textwidth]{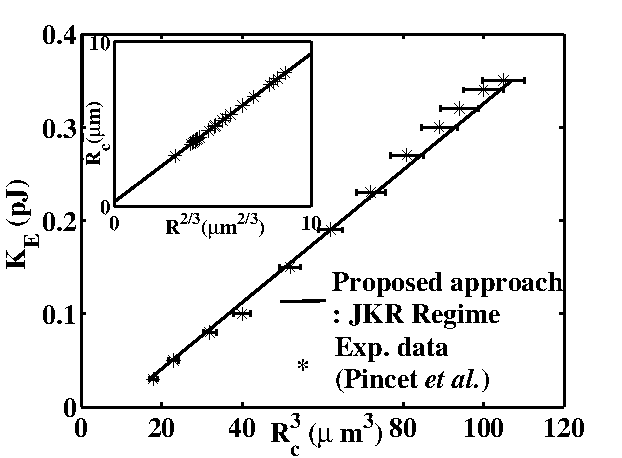}
\caption{K$_E$ vs. R$_c^3$, (Inset) R$_c$ vs. R$^{2/3}$} \label{subfig:fig2a}
\end{subfigure}
\begin{subfigure}{0.49\textwidth}
\includegraphics[width=\textwidth]{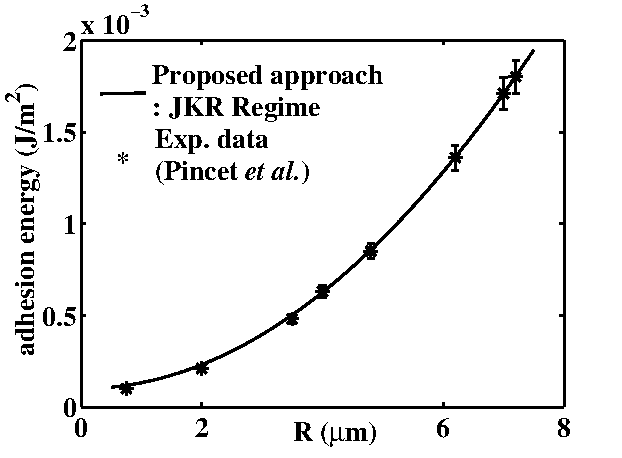}
\caption{Adhesion energy, w, vs. R} \label{subfig:fig2b}
\end{subfigure}
\caption{(\subref{subfig:fig2a}) Parameter K$_E = R_h [F_{\textrm{pull}}+3\pi R_h w + \sqrt{6\pi R_hwF_{\textrm{pull}}+(3\pi R_hw)^2}]$ versus contact radius, $R_c^3$. (Inset) contact radius versus particle radius, $R$. (\subref{subfig:fig2b}) Adhesion energy, w (Eqn.~(\ref{eqn:SurfEnergy})), versus particle radius. The experimental data in figures (\subref{subfig:fig2a}) and (\subref{subfig:fig2a}, Inset) are fitted with the proposed model in the JKR regime (Eqn.~\eqref{eqn:JKRradius}). The slope of the curve in this case gives the elastic modulus, K. In figure (\subref{subfig:fig2b}), the curve represents the adhesion energy, $w_{JKR}=-2F_{\text{pull}}/(3\pi R_h)$. The curves are fitted within a 3\% accuracy with the model parameters fixed at $\lambda=10^{-8}Nm^{-1}, \kappa=5.0$.}\label{fig:fig2}
\end{figure}

As a preliminary step, the model was validated with the experimental data reported by Pincet {\it et al.}~\citep{Pincet2005} in the JKR regime, i.e., when the deformation is substantial and the Tabor ratio is larger than one. This is achieved by fixing the model parameters at $\lambda=10^{-8}$Nm$^{-1}$ and $\kappa=5.0$. The closeness of fit between the model (highlighted by the solid lines) and the experimental data points, for variable contact radius, $R_c$, versus the elastic modulus, $K_E$, is shown in figure~\ref{fig:fig2}\subref{subfig:fig2a} and for adhesion energy ($w_{JKR}=-2F_{\text{pull}}/(3\pi R_h)$ for solid homogeneous spheres, as predicted by the JKR theory) versus the particle size, $R$, is presented in figure~\ref{fig:fig2}\subref{subfig:fig2b}, respectively. The error bars represent the maximum and minimum deviation from the sample points and set at 3\% margin of error. In figure~\ref{fig:fig2}\subref{subfig:fig2a}, the slope of the line estimates the value of the elastic modulus $K=3556$ (in Eqn.~\eqref{eqn:JKRradius}) which is in excellent agreement with reported experimental literature~\citep{Leung1984,Pincet2005}. Notice the linear scaling relationship between the contact radius and the particle size (figure~\ref{fig:fig2}\subref{subfig:fig2a} (Inset)) accurately which is captured by the model.


\subsection{Force, normal stress, contact area, adhesion energy} \label{subsec:AE}
\begin{figure}
\centering
\begin{subfigure}{0.49\textwidth}
\includegraphics[width=\textwidth]{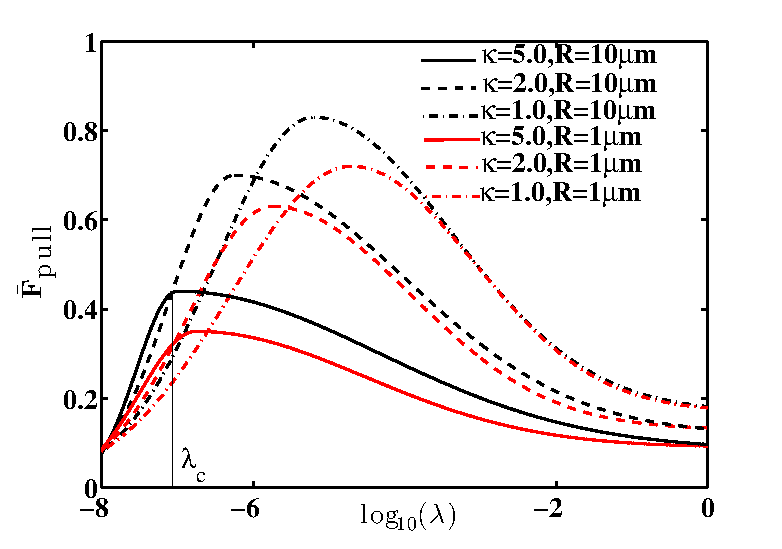}
\caption{Magnitude of the pull off force} \label{subfig:fig3a}
\end{subfigure}
\begin{subfigure}{0.49\textwidth}
\includegraphics[width=\textwidth]{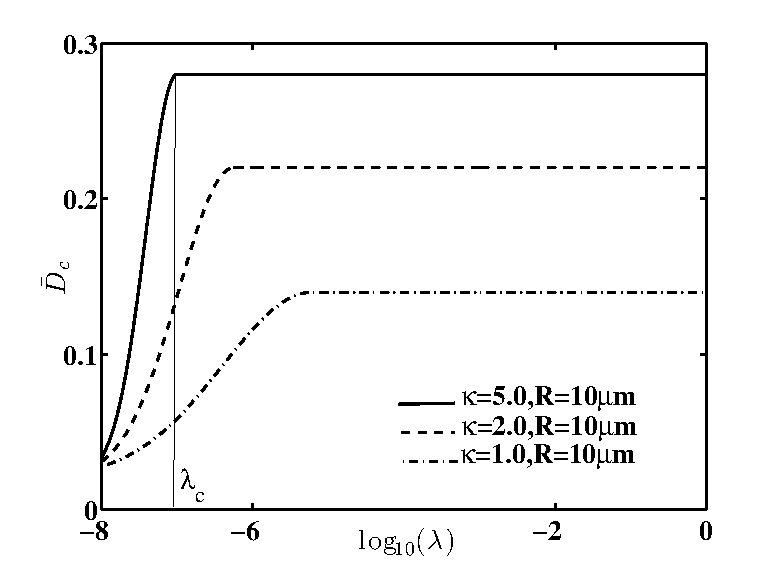}
\caption{Critical separation distance} \label{subfig:fig3b}
\end{subfigure}
\begin{subfigure}{0.49\textwidth}
\includegraphics[width=\textwidth]{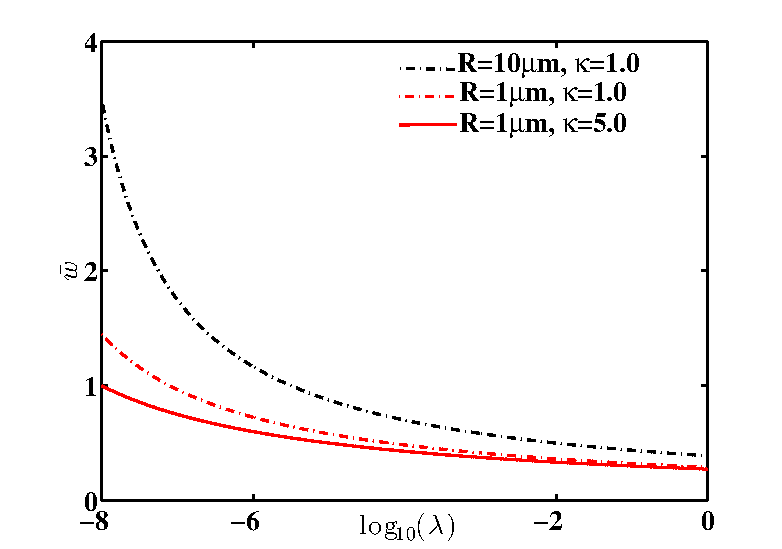}
\caption{Adhesion energy versus binder stiffness} \label{subfig:fig3c}
\end{subfigure}
\caption{Normalized (\subref{subfig:fig3a}) pull off force, $F_{\text{pull}}/F_0$, (\subref{subfig:fig3b}) critical separation distance, $\bar{D}_c=R(\zeta_c+{\it l}_0)$, where $\zeta_c$ is the critical deformation at pull off, and (\subref{subfig:fig3c}) adhesion energy, $\bar{w}=F_0R w$, versus the binder stiffness coefficient, $\lambda$.}\label{fig:fig3}
\end{figure}

\begin{figure}
\centering
\begin{subfigure}{0.49\textwidth}
\includegraphics[width=\textwidth]{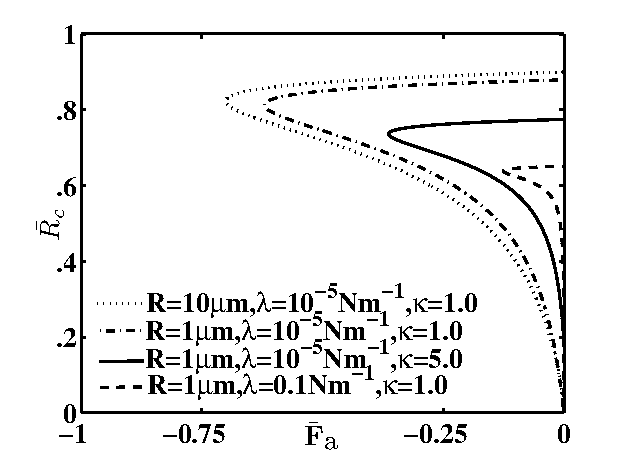}
\caption{Contact radius versus adhesive force} \label{subfig:fig4a}
\end{subfigure}
\begin{subfigure}{0.49\textwidth}
\includegraphics[width=\textwidth]{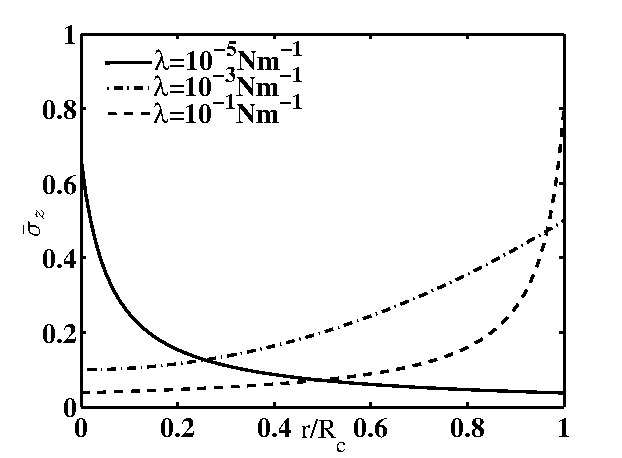}
\caption{Normalized stress at $\kappa=1.0$} \label{subfig:fig4b}
\end{subfigure}
\caption{(\subref{subfig:fig4a}) Contact radius, $\bar{R}_c=R R_c$, versus the adhesive force, $\bar{F}_a=F_0 F_a$, and (\subref{subfig:fig4b}) normal stress distribution, $\bar{\sigma}_z=\nicefrac{{\sigma}_z}{(F_0/R^2)}$, inside the contact region for particle of size $R=1\mu$m.}\label{fig:fig4}
\end{figure}

Next, we explored the binder kinematics (i.e., the magnitude of the pull off force figure~\ref{fig:fig3}\subref{subfig:fig3a}) of the elastic spheres versus the binder stiffness, $\lambda$, in different ionic conditions affecting the screening length, $\kappa$, and variable particle size, $R$. In general, strong adhesion (represented by a greater magnitude of the pull off force) is observed for larger particles and highly charged surface (i.e., shorter screening lengths,~$\kappa$). Since the area of adhesion is proportional to the size of the particle (e.g., compare the radius of adhesion curves for particles of size 10$\mu$m versus those of size 1$\mu$m in figure~\ref{fig:fig4}\subref{subfig:fig4a}), Jensen and colleagues~\citep{Hodges2002,Reboux2008} have suggested that a bigger adhesion region implies more ligands available for binding. Similarly, a shorter screening length signifies a smaller separation distance between the interacting surfaces, and hence a strong adhesion.

The nonlinear relation between the pull off force and the binder stiffness is justified as follows. At close proximity, the short range Columbic force~$\vec f_C$, pushes the particle farther away. For sufficiently large separation gap, the bonds stretch and the extension forces,~$\vec f_E$ ($\propto D_c$), tend to pull the surfaces close to each other. However, inflexible bonds (e.g., consider the pull off force curves in the range $\lambda > 10^{-4}$Nm$^{-1}$, figure~\ref{fig:fig3}\subref{subfig:fig3a}) yield and rupture quickly leading to a rapid decay in the pull off force to a state where the spherical surfaces are free from nearly all adhesive bonds. A feeble, non-zero pull off force for highly stiff binders (e.g., $\lambda > 10^{-1}$Nm$^{-1}$, figure~\ref{fig:fig3}\subref{subfig:fig3a}) and a small non-zero work of adhesion (e.g., figure~\ref{fig:fig3}\subref{subfig:fig3c}) is due to adhesion triggered by the surface charges. Thus, the pull off force depends on the critical separation gap, $\bar{D}_c=\bar{D}_c(R, \kappa, \lambda)$, (figure~\ref{fig:fig3}\subref{subfig:fig3b}), which varies nonlinearly with the binder stiffness, ionic conditions of the interacting surface and the particle size, which accounts for the non-linear variation versus the separation distance. 

Altogether, the particle size, binder elasticity and the surface charge modulates the ligand mediated adhesion kinetics. Bigger particles with highly charged surface (represented by a shorter Debye length~$\kappa$) coated with flexible tethers exhibit compact binding with a higher magnitude of adhesive force (figure~\ref{fig:fig4}\subref{subfig:fig4a}). Dooki et al.~\citep{Dooki2015} provided an alternate analogy based on the adhesion energy of binding surfaces (figure~\ref{fig:fig3}\subref{subfig:fig3c}) and again concluded that a strong adhesion (indicated by a greater work done against the adhesive force to separate the two surfaces) is an outcome of the above mentioned factors.

The normal stress distribution ($\bar{\sigma}_z=\nicefrac{\sigma_z}{(F_0/R^2)}$, figure~\ref{fig:fig4}\subref{subfig:fig4b}) revealed a peculiar picture inside the area of contact: in case of flexible binders the stress is maximum at the center of the contact region, while it is maximum at the boundary of the region when the binders are stiff. Stress distribution profile gradually varies between these two extremes as the binder stiffness is changed. These results allow for the following physical interpretation: for elastic binders, the maximum compression of the sphere surface occurs at the point of minimum separation, i.e., at the center of the contact region. For stiff binders, the problem approaches that of a rigid punch (the binders) penetrating an elastic half-space (the sphere). In the latter case, Johnson~\citep{Johnson1985} predicted that the normal stress becomes infinite at the periphery of the contact region. In another numerical experiment, we found that the normal stress reduces for weakly charged surfaces (i.e., longer screening length, $\kappa$), although the qualitative profile remains the same, indicative of the case of weak adhesion for moderately charged surfaces. The results of these experiments are not shown here for conciseness.

\subsection{Effects of surface heterogeneity} \label{subsec:SH}
Finally, we report our numerical findings on the qualitative effects of the  surface heterogeneities on the mechanics of ligand-mediated sphere adhesion. To comprehend the impact of the irregularities, we define the reduction ratio of the pull off force, $\delta_r$, as
\ben
\delta_r = 1 - \frac{F^r_{\text{pull}}}{F_{\text{pull}}}, 
\een
where $F^r_{\text{pull}}$ is the pull off force for a rough surface. 

\begin{figure}
\centering
\begin{subfigure}{0.49\textwidth}
\includegraphics[width=\textwidth]{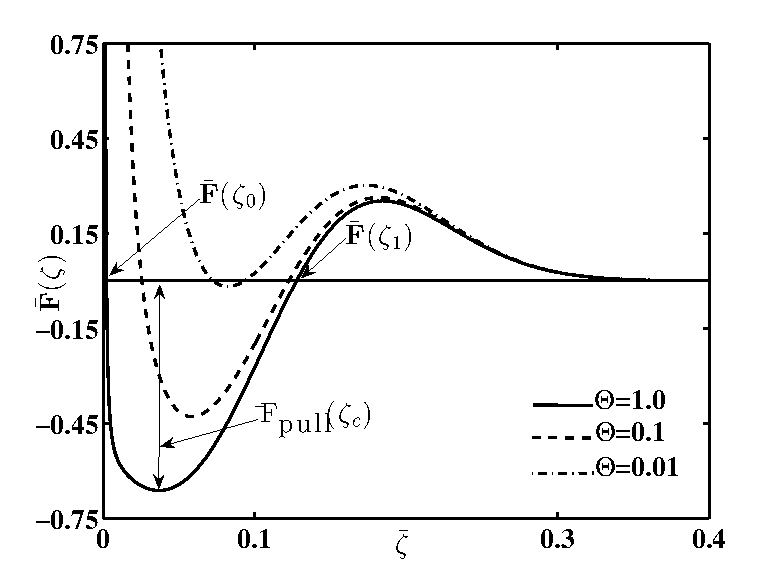}
\caption{Normalized force versus net deformation} \label{subfig:fig6a}
\end{subfigure}
\begin{subfigure}{0.49\textwidth}
\includegraphics[width=\textwidth]{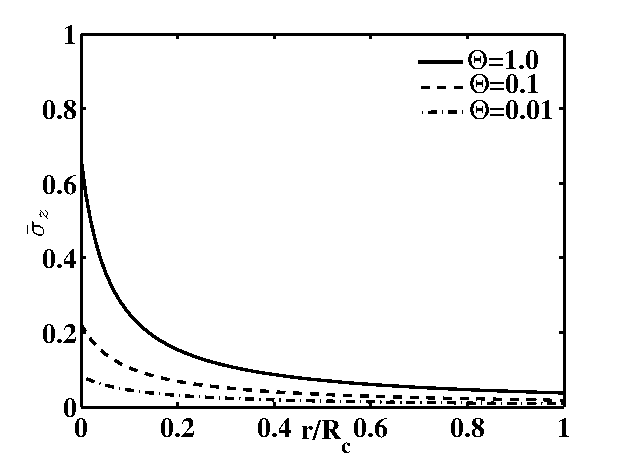}
\caption{Normalized stress distribution} \label{subfig:fig6b}
\end{subfigure}
\begin{subfigure}{0.49\textwidth}
\includegraphics[width=\textwidth]{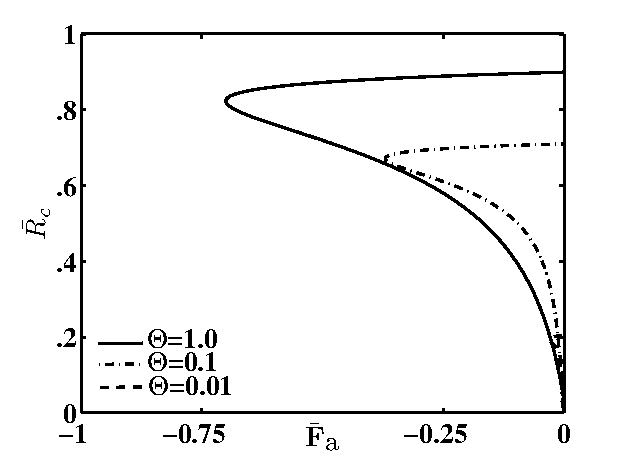}
\caption{Contact radius versus adhesive force} \label{subfig:fig6c}
\end{subfigure}
\caption{Normalized (\subref{subfig:fig6a}) force, ${\bf F}/F_0$, versus net surface deformation, and (\subref{subfig:fig6b}) normal stress distribution within the area of contact, and (\subref{subfig:fig6c}) contact radius, $\bar{R}_c=R R_c$, versus the adhesive force, $\bar{F}_a=F_0 F_a$, for particles of size $R=1\mu$m. The material parameters for these simulations is fixed at $\lambda=10^{-5}$Nm$^{-1}$, $\kappa=1.0$.}\label{fig:fig6}
\end{figure}

Surface heterogeneity significantly weakens particle adhesion. For example, for charged spherical particles coated with elastic binders (i.e., $\kappa=1.0, \lambda=10^{-5}$ Nm$^{-1}$, figure~\ref{fig:fig6}\subref{subfig:fig6a}), the pull off force reduces from $F_{\text{pull}}=-0.27$ for smooth surfaces ($\Theta=1.0$) to $F_{\text{pull}}=-0.18$ for moderately rough surface ($\Theta=0.1$), with a reduction of $\delta_r=0.33$. In the case of a highly jagged surface (($\Theta=0.01$)), this force dwindles to only $F_{\text{pull}}=-0.02$ (a reduction of 93\%).

Highly erratic surface implies that the equilibrium separation gap is higher (e.g., consider the incremental shift in the critical gap at pull off, D$_c (\propto \zeta$), figure~\ref{fig:fig6}\subref{subfig:fig6a}), while the normal stress is lower (figure~\ref{fig:fig6}\subref{subfig:fig6b}). Microparticles with irregular surface are essentially undeformed at pull off. The adhesion area is small (figure~\ref{fig:fig6}\subref{subfig:fig6c}) and this area shrinks as the surface heterogeneity increases. Conversely, smooth microparticles adhere tenaciously due to a lower gap width, resulting in a strong normal stress and a robust attractive force--
a phenomena which has observed in a variety of {\it in vivo} experiments by Toika et al.~\citep{Toika2000}, later corroborated by Brach et al.~\citep{Brach2002}.

\section{Conclusions} \label{sec:end}
Section~\S \ref{sec:model} presented a comprehensive, unified, multi-scale model for the adhesion mechanics of two static, heterogeneously charged and deformable, solid, spherical microparticles coated with binding ligands. Section~\S \ref{sec:results} validates the model in the regime of application of the JKR theory and underlines the role of particle size, surface charge, binder elasticity and surface heterogeneity on the adhesion strength of the coalescing particles. In particular, smooth, highly charged surface, covered with flexible binders exhibit strong adhesion, indicated by a substantial increase in the pull off force, closer separation, strong normal stress, higher adhesion energy and enlarged area of adhesion. These observations have been corroborated in numerous {\it in vivo} studies on ligand-mediated bioadhesion~\citep{Hammer1996, Zhang2013}.

The proposed model is a preliminary step to describe key features in ligand mediated surface adhesion, coupled with the mechanics of deformed surface, hence several issues still need to be addressed. For example, the nonlinearity of the micro-scale forces can induce both tangential and an out of plane torsional shear~\citep{Johnson1997} thereby modifying the area of adhesion. Our approach also excludes spatial inhomogeneity arising through the material parameters, the effects of catch behavior ($\lambda_s > \lambda$), non-equilibrium binding effects, stochasticity and the discrete number of bonds~\citep{Zhu2000}, material viscoelasticity (needed to fully describe the particle rheology~\citep{Biggs2000}) as well as shearing forces large enough to tear the binding ligands from their anchoring surface~\citep{Varenberg2007}. All these effects can lead to several non-trivial behavior (including the possible presence of a hysteresis in contact radius-adhesive force curves) that deserves a full numerical investigation in the near future.

\section*{Acknowledgements}
SS acknowledges the financial support from Dr. David Bortz in Dept. of Applied Mathematics, University of Colorado, Boulder, where the work started initially. Further, financial support of the Adelaide University startup funds and the Australian Research Council Discovery grant DP150102385, are gratefully acknowledged.


\end{document}